\documentclass[fleqn,12pt,twoside]{article}
\usepackage{espcrc1,epsf}
\usepackage{graphicx}
\usepackage[figuresright]{rotating}
\title{GROUND STATE CORRELATIONS AND FINAL STATE INTERACTIONS IN TWO-NUCLEON EMISSION PROCESSES
OFF FEW-NUCLEON SYSTEMS\footnote{Presented at the {\it 5-th
Workshop on "e-m induced Two-Hadron Emission"}, Lund,
June, 2001.}}

\author{C. Ciofi degli Atti\address{ Department of
 Physics, University of Perugia, and
 INFN, Sezione di Perugia,
 via A. Pascoli, Perugia, I-06100, Italy}
   \, and     L.P. Kaptari$\, ^{\rm a}$
\thanks{On leave from  Bogoliubov Lab. Theor. Phys.,
        JINR, Dubna, Russia} }
\begin{document}

\def \bfgr #1{ \mbox {{\boldmath $#1$}}}
\newcommand{\be}{\begin{eqnarray}}
\newcommand{\ee}{\end{eqnarray}}
\newcommand{\nn}{^3He(e,e'N_2N_3)N_1}
\newcommand{\pp}{^3He(e,e'p_1p_2)n}
\newcommand{\pn}{^3He(e,e'p_1n)p_2}
\newcommand{\ga}{\gamma^*}
\newcommand{\emme}{M({\bf p}_m,{\bf t})}
\newcommand{\spectral}{P_1(k_1,E^*)}
 \maketitle

\begin{abstract}
  The two nucleon emission process from $^3He$,
  $^3He(e,e'N_2N_3)N_1$, has been theoretically analyzed using realistic
  three-nucleon wave functions and taking the final state interaction into account.
  Various kinematical conditions have been considered in order to
  clarify the question  whether the effects
  of  the final state interaction could be  minimized by a proper choice of the
  kinematics of the process.
\end{abstract}
\section{GROUND STATE CORRELATIONS IN NUCLEI}
The  investigation of Ground State  Correlations (GSC) in nuclei, in particular those which
originate from the most  peculiar features of the Nucleon-Nucleon (NN)  interaction,
i.e. its strong short range repulsion  and  complex state  dependence (spin,
isospin, tensor, etc),
is one of  the most challenging aspects of experimental and theoretical
nuclear physics and, more generally, of hadronic physics. The results of sophisticated
many-body calculations in terms of realistic models of the NN interactions, show
 that the complex structure of the latter generates a rich correlation structure of the
 nuclear ground state wave function.
The experimental investigation of the nuclear wave function or, better, of various
density matrices, $\rho (1), \rho(1,1'), \rho(1,2), etc$ , is therefore necessary in order to
 ascertain whether the prediction of the  {\it Standard Model} of nuclei (structureless non-
 relativistic nucleons interacting via the free NN interaction) is indeed justified in practice, or
 other phenomena, e.g.:
\begin{enumerate}
\item effective single-particle mean field,
\item relativistic effects,
\item many-body forces,
\item medium modification of nucleon properties,
\item explicit sub-nucleonic degrees of freedom (quark and gluons),
\end{enumerate}
have to be advocated in order to describe ground-state properties of nuclei at normal
density and temperature.

Unfortunately,
whereas the one-body density matrix (charge density)  is experimentally well known since many
years (see e.g.\cite {report}),
the present knowledge of those quantities, e.g. the non-diagonal one-body (momentum distributions)
and  two-body (two-nucleon correlation function) density matrices, which could
provide more reliable  information on GSC, is still too  scarce. One the reasons for that
should be ascribed
to the effects of the final state interaction (FSI),  which very often compete  with  the
effects generated by GSC. The present situation is therefore such that
 the longstanding question:
 \begin{itemize}
  \item {\it Does FSI hinder the investigation of GSC?}
  \end{itemize}
  has not yet been
 clearly answered. Moreover, due to the difficulty to treat consistently GSC and FSI  within a many-body
  approach,
 the answer to the above
  question was in the past merely dictated by philosophical taste  rather than by the results
 of solid calculations and by unambiguous experimental data. To-day the answer could  probably  be
  provided
 in a more reliable way, particularly in the case of few-body systems, for which accurate
 ground state wave functions are available and  FSI effects can also be calculated in a
 satisfactory way (see e.g. \cite{gloeckle}, \cite{rosati}). In this paper we will discuss
 the process of
  two-nucleon emission off the three-nucleon systems, which is being experimentally investigated
 (see e.g. \cite{larry}, \cite{eddy}), with the aim of providing
 another attempt  at answering
  the above longstanding question.
 The results of some theoretical
 calculations,  using realistic three-nucleon wave function, and taking into account
 FSI effects will be presented, and future perspectives in the field briefly discussed.

 \section{Two-nucleon emission off the three-nucleon systems:
 kinematics and cross section}
 We will consider the absorption of a virtual photon ${\gamma^*}$ by  a nucleon bound in $^3He$, followed
 by two-nucleon emission. In the rest of this paper
 the photon four momentum transfer will be denoted by  $Q^2={\bf q}^2-{\nu}^2$,
 the  momenta of the bound nucleons, before ${\gamma^*}$ absorption,  by $\bf k_i$,
 and the momenta in the continuum
 final state by $\bf p_i$.
\begin{figure}[h] 
\begin{center}
\includegraphics[height=0.20\textheight]{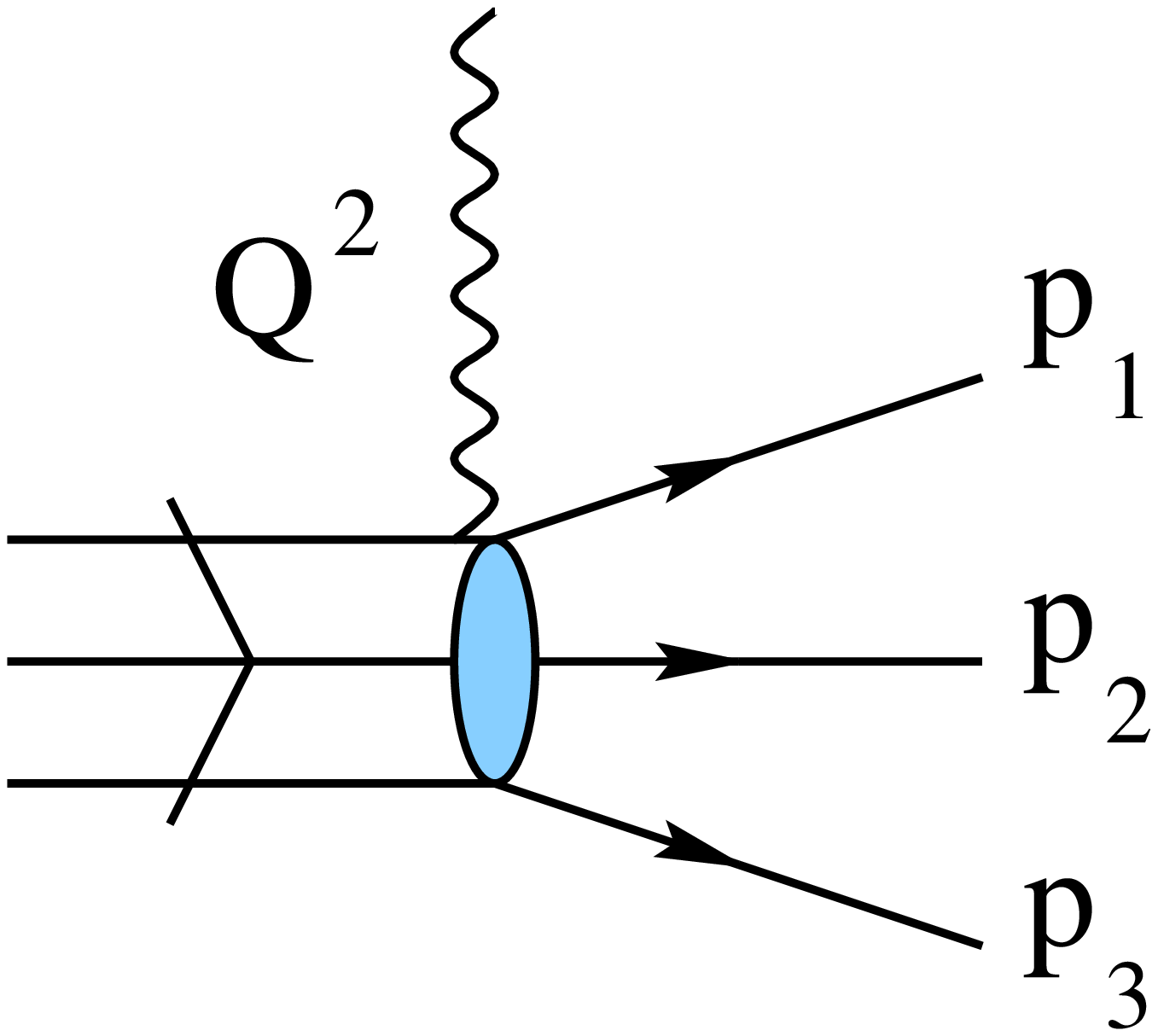}
\hfill
\begin{minipage}{6in}
\caption{The One Photon Exchange diagram for the
    process $\nn$}—
    \label{fig1}
    \end{minipage}
\end{center}
\end{figure}
 Momentum conservation requires that

 \begin{equation}
 {\sum_{i=1}^3{\bf k}_i} =0 \qquad\qquad  {\sum_{i=1}^3{\bf p}_i = {\bf q}}
 \label{one}
 \end{equation}
and energy conservation that
\begin{equation}
  \nu +M_3 =\sum_{i=1}^3 (M^2+{\bf p}_i^2)^{1/2}
  \label{two}
  \end{equation}
 where $M$ and $M_3$ are the nucleon and the three-nucleon system masses, respectively.

In what follows, the two nucleons which are detected will be denoted by $N_2$ and $N_3$ and
the third one by $N_1$. In one-photon exchange approximation the cross section of the
process, depicted in Fig. \ref{fig1}, reads as follows
\begin{equation}
\frac{d^{12}\sigma}{d\epsilon^{'} d\Omega{'} d{\bf p_1} d{\bf p_2} d{\bf p_3}}={\sigma}_{Mott}\cdot
\sum_{i=1}^6 v_i \cdot W_i \cdot \delta( {\bf q} - \sum_{i=1}^3{\bf p}_i ) \delta( \nu +M_3 -\sum_{i=1}^3 (M^2+{\bf p}_i^2)^{1/2})
\label{three}
\end{equation}
where $v_i$ are well known kinematical factors,  and $W_i$  the {\it response functions},
which have the following general form
\be
W_i \propto\left | \langle \Psi_f^{(-)}({\bf p}_1, {\bf p}_2, {\bf p}_3|\hat{ \mathcal O}_i({\bf q})|
\Psi_i({\bf k}_1, {\bf k}_2, {\bf k}_3\rangle \right |^2
\label{four}
\ee
In Eq.\ref{four} $|\Psi_f^{(-)}({\bf p}_1, {\bf p}_2, {\bf p}_3\rangle$ and
$|\Psi_i({\bf k}_1, {\bf k}_2, {\bf k}_3\rangle$
are the continuum and ground state wave functions of the three body system, respectively,
and   $\hat{ \mathcal O}_i({\bf q})$ is a  quantity depending on proper combinations of the components
of the  nucleon
current  operator  $\hat{ j^{\mu}}$ (see e.g. \cite{report}).
 If  FSI is fully disregarded, two nucleon emission originated by $NN$ correlations can occur because
 of two
different processes:
\begin{enumerate}
\item in the initial state $N_2$ and $N_3$ are correlated and $N_1$ is far apart; $\gamma^*$ is
absorbed by $N_1$ and $N_2$ and $N_3$ are emitted in the continuum;
\item in the initial state $N_1$ is correlated with either $N_2$ or $N_3$ (e.g. $N_2$);
$\gamma^*$ is absorbed by $N_1$   and  $N_2$ and  $N_3$   are emitted in the
continuum.
\end{enumerate}
In both cases momentum conservation reads as follows
\be
{\bf p}_1= {\bf k}_1 +{\bf q} \qquad\qquad {\bf p}_2= {\bf k}_2\qquad \qquad  {\bf p}_3= {\bf k}_3
\label{five}
\ee
Thus, if the uncorrelated nucleon was at rest in the initial state, in case $a)$
the correlated nucleons $N_2$ and $N_3$ are emitted back-to-back, and in case
$b)$ $N_2$ is emitted with momentum ${\bf p}_2= {\bf q}- {\bf p}_1$ and $N_3$
with momentum  ${\bf p}_3=0$.

The above picture is distorted by the final state interaction . In what follows we will
investigate  process $1.$, in particular we will investigate how
FSI will distort the simple picture described above.
\section{The  Final State Interaction in the process
${^3H\lowercase{(e,e'}N_2N_3)N_1}$}

\begin{figure}[h] 
\begin{center}
    \includegraphics[height=0.50\textheight]{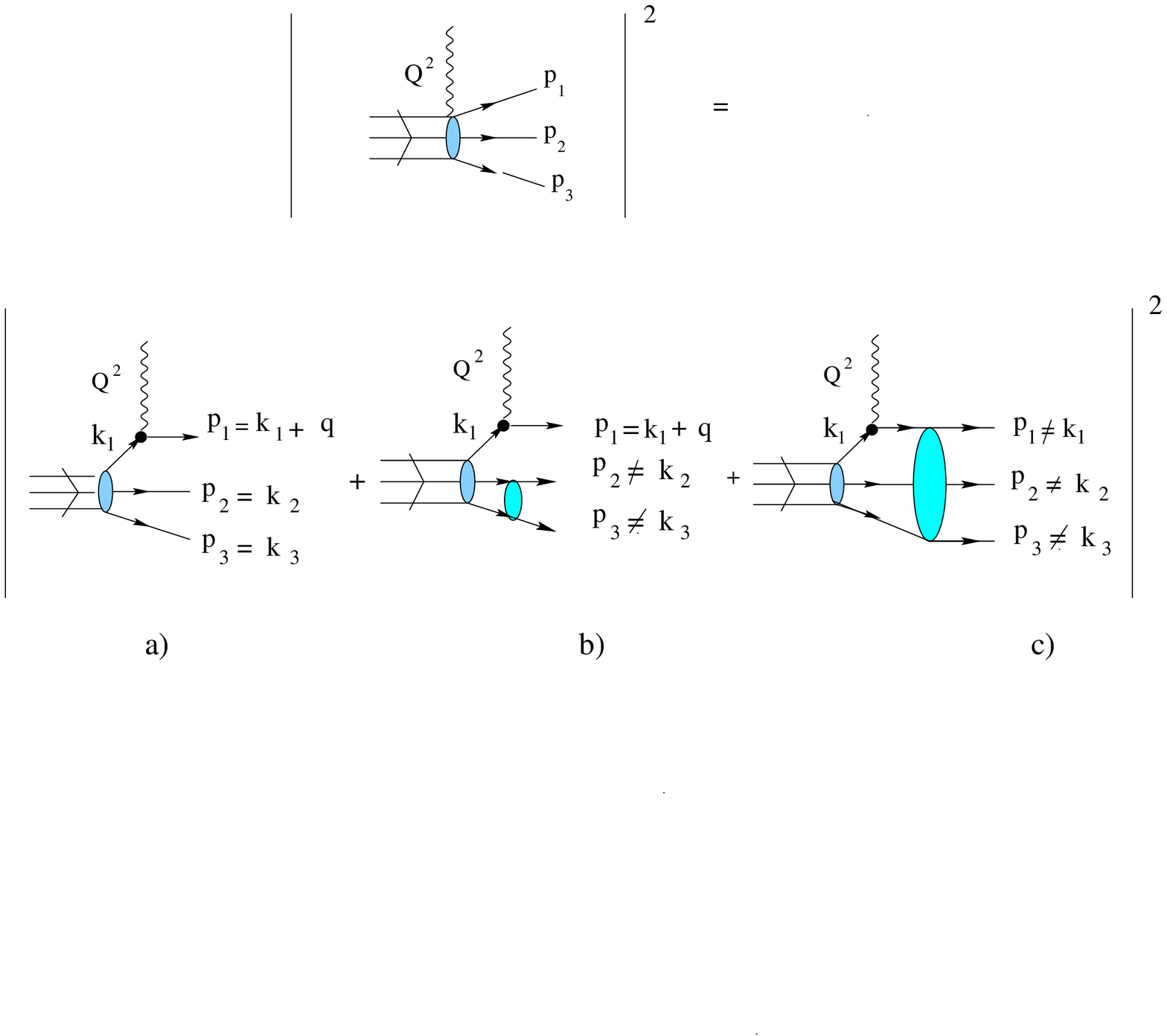}
\vskip -4cm
\caption{ The various processes contributing to  the
 reaction  $\nn$: (a) will be called {\it No FSI}, (b) will be
 called {\it the NN rescattering}, (c) will be called {\it the three-body
 rescattering}.
 Note that in this paper, following Ref. \protect\cite{CPS},
  the sum of $a)$ and $b)$ is called
 {\it The Plane Wave
 Impulse Approximation (PWIA)}, whereas in Ref. \protect\cite{gloeckle}
  PWIA is the same as our {\it No FSI}
 (process $b)$).}
\label{fig2}
\end{center}
\end{figure}

 The various processes, in order of increasing complexity,
 which contribute to the reaction $^3He(e,e'N_2N_3)N_1$ are depicted in
 Fig. \ref{fig2}.

  Let us introduce  the following quantities:
 \begin{enumerate}
 \item the {\em relative momentum} of the detected
 pair
 \be
 {\bf t}=\frac{{\bf p}_2-{\bf p}_3}{2}
 \label{six}
 \ee
 \item the {\em Center-of-Mass momentum}  of the pair
 \be
 {\bf P}={\bf p}_2+{\bf p}_3
 \label{seven}
 \ee
 \item  the {\it missing momentum}
 \be
{\bf p}_m={\bf p}_1-{\bf q}= -({\bf p}_2+{\bf p}_3)
\label{eight}
\ee
\end{enumerate}

As already stated, we will  consider the process  $\pp$ ( $\pn$) , in which $\ga$
 interacts with the neutron (proton) and the two protons (proton-neutron) {\it correlated in the
 initial state} are emitted and detected. Let us  disregard, for the time being,
 the interaction of the hit neutron(proton) with the emitted proton-proton
 (proton-neutron) pair, but take into account the final state rescattering
 between the two detected nucleons.
 Then the processes contributing to the cross section are $a)$ and $b)$ of  Fig. \ref{fig2}.
By integrating the  cross section (Eq. \ref{three})
over ${\bf P}$ and the kinetic energy of $N_1$, and taking  ${\bf q }\parallel  z$,  one obtains
\be
\frac{d^{8}\sigma}{d\epsilon^{'} d\Omega{'} d\Omega_{N_1} d{t}  d\Omega_t}
={\mathcal K}\cdot M({\bf p_m}, {\bf t})
\label{nine}
\ee
where  ${\bf p}_m= - {\bf k}_1$, ${\mathcal K}$ incorporates all
kinematical factors, and
\be
M({\bf p}_m, {\bf t})= M({\bf k}_1, {\bf t})=\frac12
\sum\limits_{M_3,\sigma}\sum\limits_{s_f,\mu_f}
\left|\int \exp( {i {\bf p}_m \bfgr\rho})
\, \chi_{\frac12 \,\sigma}
\Phi_{s_f\mu_f}^{{\bf t}(-)}({\bf r}) \Psi_{3M_3}^*({\bf r},{\bfgr \rho})\right |^2
\label{ten_a}
\ee
where $\chi_{\frac12 \,\sigma}$ represents the Pauli spinor for the hit particle,
$\Phi_{s_f\mu_f}^{{\bf t}(-)}(\bf r)$ is the two-nucleon wave function in the continuum, i.e.

\be
\Phi_{s_f\mu_f}^{{\bf t}(-)}({\bf r})=\sum\limits_{l,m,\lambda}\sum\limits_{J(M_J)}
\langle lms_f\mu_f|JM_J\rangle\, Y_{lm}(\hat {\bf t})\,R^{J\, |{\bf t}|}_{ls_f\lambda}(r)
\Upsilon_{ l'(\lambda)s_f}^{JM_J}( \hat{\bf r}) \, {\cal T}_f
\label{ten_b}
\ee
 $ {\cal T}_f$ is the isospin function of the final pair, and the other notations
are self-explaining. In the rest of the paper we will omit, for ease of presentation,
all explicit summations over the quantum numbers and will denote the
continuum two-nucleon wave function simply by  $\Phi_{N_2N_3}^{{\bf t}^{(-)}}$.
 Equation   (\ref{ten_a}) can then be cast in the following simple
form
\be
M({\bf p}_m, {\bf t})= M({\bf k}_1, {\bf t})=  \left|\int \exp( {i {\bf p}_m \bfgr\rho})
\, I_{N_2N_3}^{\bf t}({\bfgr \rho})d^3\rho \right|^2
\label{ten}
\ee
where   $I_{N_2N_3}^{\bf t}({\bfgr \rho})$ is the {\it overlap  integral}
between the three-nucleon ground state wave function and the two-nucleon continuum state,
 i.e.
\be
 I_{N_2N_3}^{{\bf t}}({\bfgr \rho})=\int\Phi_{N_2N_3}^{{\bf t}^{(-)}} ({\bf r})\Psi_{3M_3}({\bf r},
  {\bfgr \rho})d^3 r
\label{eleven}
\ee
where ${\bf r}$ and  $\bfgr \rho$ are usual Jacobi coordinates.

If also the NN rescattering is disregarded  ( {\it No FSI} approximation (process $a)$),  one has
\be
\Phi_{N_2N_3}^{{\bf t}^{(-)}}\propto e^{i{\bf t} \cdot {\bf r}}
\label{twelve}
\ee
\be
 I_{N_2N_3}^{{\bf t}}({\bfgr \rho})=\int e^{i {\bf t}{ \bf r}}
 \Psi_{3M_3}({\bf r}, {\bfgr \rho})d^3 r
\label{3teen}
\ee
and
\be
M({\bf k}_1, {\bf t})=\left| \int \exp( {i {\bf k}_1 {\bfgr {\rho}}}) \exp({i {\bf t}{ \bf r}})
\Psi_{3M_3}({\bf r},
 {\bfgr \rho})d^3 r
d^3\rho \right|^2
\label{4teen}
\ee
which represents nothing but the three-nucleon wave function in momentum space. Thus, if FSI
is fully disregarded, the process $\nn$
would be directly related to the three-nucleon
wave function in momentum space.

\begin{figure}[t] 
\epsfxsize 4in
\centerline{ \epsfbox{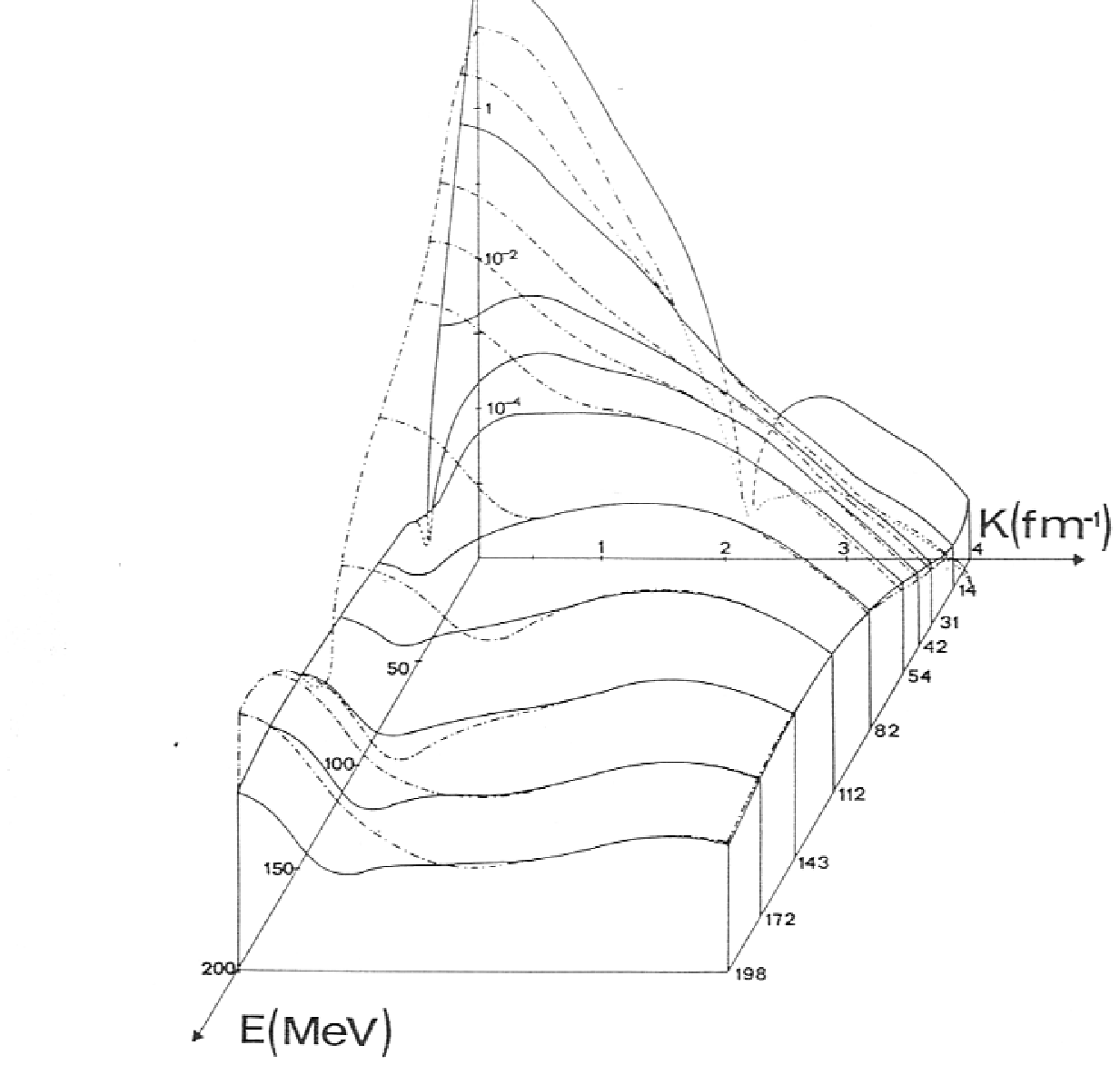}}
\caption{
The three-body channel neutron (proton) Spectral Function
in $^3He$ ($^3H$) ($k\equiv k_1 $). The dot-dashed line
corresponds to the {\it No FSI} case, whereas the full line includes the
 neutron-neutron (proton-proton)  rescattering. Three-nucleon
 wave function from \protect\cite{CPS}; Reid Soft Core interaction
\protect\cite{Reid}
(After Ref. \protect\cite{CPS}).
}
\label{fig3}
\end{figure}

Let us now switch on  the NN rescattering (process $a)$ plus process
$b)$, i.e. the PWIA). To this end it is worth pointing out  that if the overlap integral is integrated over the direction of ${\bf t}$, the nucleon
($N_1$) {\it Spectral Function} is obtained, {\it viz}
\be
\int M({\bf k}_1, {\bf t}) d\Omega_{\bf t}=P_1(k_1,E^*)
\label{5teen}
\ee
where
\be
 E^*=\frac{{\bf t}^2}{M}
 \label{6teen}
 \ee
 is the "excitation energy" of the spectator pair $N_2N_3$, which is related to the {\it removal energy} $E$
 of nucleon $N_1$ by
 \be
 E= E_3 + E^*
 \label{7teen}
 \ee
 where $E_3$ is the (positive) binding energy of the three nucleon system.

Thus, by integrating Eq. \ref{nine} over $\Omega_t$, one gets
\be
\frac{d^{6}\sigma}{d\epsilon^{'} d\Omega{'} d\Omega_{N_1} d{t}}
= \mathcal K\cdot P(k_1,E^*)
\label{8teen}
\ee

\begin{figure}[h] 
 \vskip .7cm
\begin{center}
    \includegraphics[height=0.40\textheight]{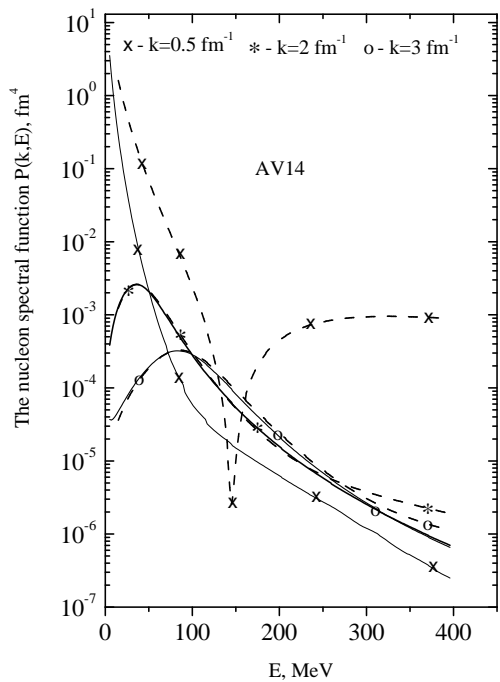}
\caption{The Nucleon Spectral Function as in Fig.\protect\ref{fig3}
corresponding to the wave function
of Ref. \protect\cite{rosati}
and the $AV14$ interaction \protect\cite{AV14} ($k\equiv k_1$).
 The dashed line
corresponds to the {\it No FSI} case, whereas the full line includes the
 neutron-neutron (proton-proton)  rescattering
 (process $b)$ in Fig. \protect\ref{fig2}.}
\label{fig4}
\end{center}
\end{figure}

If the Coulomb
 interaction is disregarded, the Spectral Function for the three-body channel
 is the same for both processes we are considering  ($\pp$ and $\pn$). Therefore,
 in order to illustrate the effect of NN rescattering, we show in Fig. \ref{fig3} the nucleon
 Spectral Function calculated with and without the NN rescattering \cite{CPS}.
  It can be seen that
 there is a  region where the  FSI  does not play any role. This is the so called
 {\it  two-nucleon correlation} region, where the relation

  \be
  E^* = \frac {{\bf t}^2}{M} \simeq \frac{{{\bf k}_1}^2}{4M}
  \label{9teen}
  \ee
 holds (see e.g. Ref. \cite{2NC}). The existence  of such a region is a general feature of any
 Spectral Function, independently of the two-nucleon interaction and of  the method to generate the
 wave function. This is illustrated in Fig. \ref{fig4} where the Spectral Function  corresponding
 to the variational wave function of Ref. \cite{rosati}and the $AV14$ \cite{AV14} interaction, is shown for
 several  values of $k \equiv k_1$.

 From the figures we have exhibited  one expects that if the kinematics is properly chosen
 (i.e. $E^* = \frac {{\bf t}^2}{M} \simeq \frac{{{\bf k}_1}^2}{4M}$) the NN rescattering can be fully
 eliminated;  on the contrary, if  it is chosen improperly (in particular corresponding to
 an initial state characterized by
  $k_1 \simeq 0$),
  the two
 nucleon rescattering fully distorts the {\it No FSI}  predictions.

\begin{figure}[h] 
  \begin{center}
    \includegraphics[height=0.550\textheight]{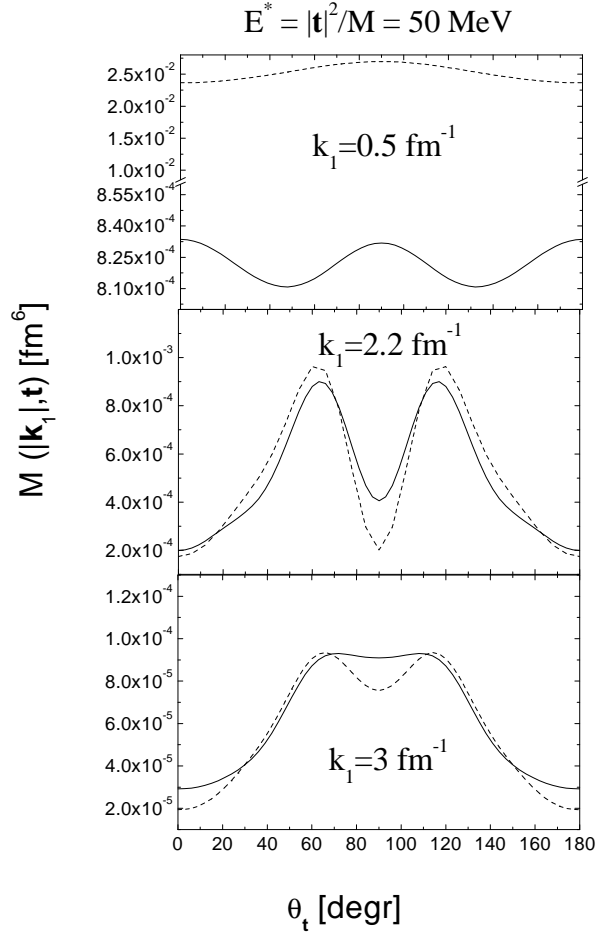}
\caption{The quantity $\emme$ (\ref{ten}) (${\bf p}_m = {\bf k}_1$)
calculated at fixed value of $E^*=50\ MeV$,
versus the angle $\theta_t$ between the relative momentum
of the emitted nucleons (\protect\ref{six})
and the momentum transfer ${\bf q}$. The full line
 includes the two nucleon rescattering and the dashed line
 represents the {\it No FSI} result.
 The three values of $k_1$ which have been chosen, correspond to
 three different regions of the Spectral Function (see text). Three-nucleon
 wave function from \protect\cite{rosati}; AV14 interaction.}
\label{fig5}
\end{center}
\end{figure}
 Since the Spectral Function is related to the cross section,
 the same effects are expected to occur on the latter.  This is indeed the case, as
  demonstrated in
  Fig. \ref{fig5}, where Eq.\ref{ten} is shown and
  compared with the {\it No FSI} approximation (Eq. \ref{4teen}).

 In this calculation,
  we have fixed the two-nucleon relative energy
  $E^*=t^2/M=50 MeV$ and have plotted, for a  given values of  $k_1$,
   the dependence of $\emme$ upon the angle $\theta_{\bf t}$ between $\bf t$ and $\bf q$,
   the latter being  chosen along
   ${\bf k}_1$. The three values of $k_1$  correspond to  three relevant regions of the
   Spectral Function, {\it viz}
   \begin{enumerate}
   \item $E^* >  \frac{{\bf k}_1^2}{4M}$ (${k}_1=0.5 fm^{-1}$);
   \item  $E^* \simeq \frac{{\bf k}_1^2}{4M}$ (${k}_1=2.2 fm^{-1}$), the correlation region;
\item  $E^* <  \frac{{\bf k}_1^2}{4M}$ (${k}_1=3 fm^{-1}$).
\end{enumerate}

\begin{figure}[h] 
  \begin{center}
  \vskip 1cm
    \includegraphics[height=0.50\textheight]{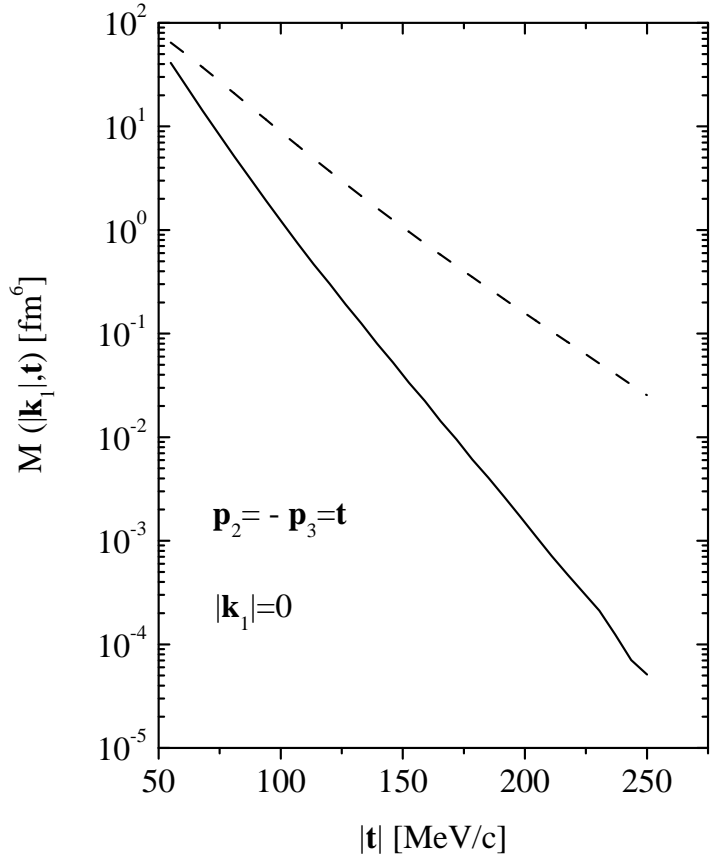}
\caption{The quantity $\emme$ (Eq. (\protect\ref{ten}))  calculated in
the kinematics where, in the initial state, nucleons $N_2$
and $N_3$ were correlated
 with momenta
 ${\bf k}_2=-{\bf k}_3$ and ${\bf k}_1=0$.
 The dashed line corresponds
 to the {\it No FSI} case, whereas
 the full line includes the two nucleon rescattering.
 Three-nucleon
 wave function from \protect\cite{rosati}; AV14 interaction.}
\label{fig6}
\end{center}
\end{figure}

It can be seen that in the first region the two nucleon rescattering is very large (cf. Figs.
\ref{fig3} and \ref{fig4}), whereas in the two other regions, it is very small.

We have also investigated the effect of $NN$ rescattering on a particular kinematics, namely that
which corresponds to the initial state in which $N_2$ and $N_3$ are correlated with momenta
${\bf k}_2$ = -${\bf k}_3$  and  ${\bf k}_1$ = $0$, so that,
 after $\ga$ absorption,
 $N_1$ is emitted  with
momentum $\bf q$  and  $N_2$ and $N_3$ are emitted  back-to-back  with momenta
${\bf p}_2$ = -${\bf p}_3$ (${\bf p}_m = 0$).

The results are presented in Fig. \ref{fig6},  where  it can be seen that,
 as expected, the effect of
 the
two-nucleon rescattering is large. We have repeated this calculation in the correlated
region and found, obviously, that the rescattering, in this case,
 has negligible effects.

\begin{figure}[h]                                      
 \vskip 1cm
  \begin{center}
    \includegraphics[height=0.40\textheight]{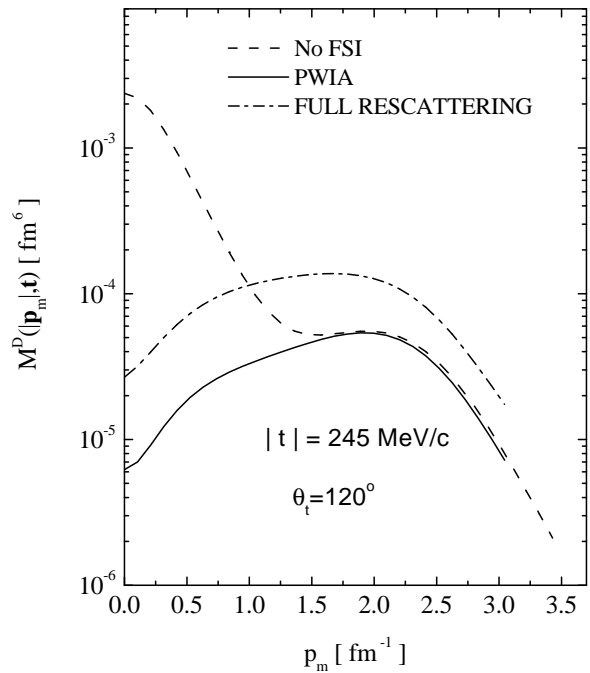}
    \caption{\label{fig7}
     The quantity $M^D({\bf p}_m, {\bf t})$,
      Eq. (\protect\ref{twenty}) (${\bf p}_m \neq {\bf k}_1$),
     calculated at fixed value of the relative
 momentum $|{\bf t}|=\, 245\,MeV/c$,
versus the missing momentum $p_m$. The full line
 includes the two nucleon rescattering,  the dashed line
 represents the {\it No FSI} result, whereas the dot-dashed line represents
 the full rescattering result. Three-nucleon
 wave function from \protect\cite{rosati}; AV14 interaction).
      }
    \end{center}
\end{figure}

We have eventually considered the three-body rescattering, e.g. process $c)$ of Fig. \ref{fig2}, by
treating the rescattering of $N_1$ with the interacting pair  $N_2N_3$ within an extended Glauber-type
approach \cite{glauber}. The details of the calculation will be presented elsewhere \cite{leonya}.
In Fig. \ref{fig7} we show ({\it preliminary results}) the
 quantity (dot-dashed line)

\be
M^D({\bf p}_m, {\bf t})=\left|\int \Phi_{N_1N_2N_3}^{{\bf p}_m} ({\bf r},{\bfgr \rho})
\, I_{N_2N_3}^{\bf t}( {\bfgr \rho}) d^3r d^3\rho \right|^2
 \label{twenty}
 \ee

  which is the generalization of Eq. \ref{ten} to take into account
   the rescattering of $N_1$ with the
 interacting pair ${N_1N_2}$, trough the quantity
 $\Phi_{N_1N_2N_3}^{{\bf p}_m} ({\bf r},{\bfgr \rho}) $.

 In the figure, $M^D({\bf p}_m, {\bf t})$ is plotted {\it vs} the
 missing momentum ( ${\bf p}_m \neq {\bf k}_1$ ) for a fixed value of ${\bf t}$;
 in the same Figure we also show   the results
 corresponding to the case when only the NN rescattering is active (full line) and to the case when
 all FSI is switched off (dashed line).  The kinematical variables
  were chosen such as to
 emphasize the correlation region of the Spectral Function. Moreover, we have considered
 high values of $|{\bf q}|$, such that the asymptotic values of those quantities which enter
 the calculation (e.g. the total NN cross section, the ratio of the imaginary to the real
 parts of the forward scattering amplitude, etc.) have been adopted.
   It can be seen that at high values of the missing
 momentum, the full FSI merely reduces to a change of the amplitude, without noticeably
   distorting
  the missing momentum distributions calculated taking
  into account only the NN rescattering contribution;
 since the latter can be reliably calculated with any realistic wave function and NN interaction,
 our result appears to be a very promising way to investigate the correlated part of the three-body
 wave function.

 \section{Conclusions}
 We have investigated the effects of the Final State Interaction in the process $\nn$ using
 realistic  three nucleon wave functions  which, being the exact solution of the
 Schroedinger equation, incorporate all types of correlations, in particular the short-range ones,
 generated by modern NN potentials. We have found that if the kinematics is chosen such as
 to emphasize the two-nucleon correlation sector of the wave function, Final State Interaction
 effects can be minimized.

  In this paper, we have not discussed other effects  competing with the Final State
 Interaction (Meson Exchange Currents, etc), which have to be analyzed before comparing theoretical
 results with experimental data.

\end{document}